\documentclass[10pt, journal]{IEEEtran}

\usepackage[OT1]{fontenc}
\usepackage{graphicx, bm, cite,array}
\usepackage{amssymb, dsfont, color, epstopdf}
\usepackage[cmex10]{amsmath}
\usepackage{enumerate}
\usepackage{apptools}
\usepackage{multirow}
\usepackage{booktabs}
\usepackage{threeparttable}
\usepackage{subeqnarray}
\usepackage{cases, stmaryrd}
\usepackage{stackengine}

\newcommand{\overbar}[1]{\mkern 1.5mu\overline{\mkern-1.5mu#1\mkern-1.5mu}\mkern 1.5mu}

\newtheorem{assumption}{Assumption}
\newtheorem{theorem}{Theorem}

\newtheorem{proposition}{Proposition}
\newtheorem{definition}{Definition}

\newcommand{\tr}{\textmd{tr}}

\newcommand{\inte}{\textrm{Int}}

\newcommand{\pr}{\mathbf{P}}
\newcommand{\ep}{\mathbf{E}}

\begin{document}

\title{\LARGE \bf Safe Stabilization for Stochastic Time-Delay Systems
\thanks{This work was supported by the Fundamental Research Funds for the Central Universities under Grant DUT22RT(3)090, and the National Natural Science Foundation of China under Grants 61890920 and 61890921.}
}

\author{Zhuo-Rui Pan, Wei Ren, and Xi-Ming Sun
\thanks{Z.-R. Pan, W. Ren and X.-M. Sun are with the Key Laboratory of Intelligent Control and Optimization for Industrial Equipment of Ministry of Education, Dalian University of Technology, Dalian 116024, China. 
Email: \textrm{\small panzhuorui@mail.dlut.edu.cn, wei.ren@dlut.edu.cn, sunxm@dlut.edu.cn}.}
}

\maketitle

\begin{abstract}
This paper addresses the safe stabilization problem of stochastic nonlinear time-delay systems. Based on the Krasovskii approach, we first propose a stochastic control Lyapunov-Krasovskii functional to guarantee the stabilization objective and a stochastic control barrier-Krasovskii functional to ensure the safety objective. Both functionals are developed respectively for each control objectives for the first time. Since the optimization problem is not easy to be resolved for stochastic time-delay systems, we derive a sliding mode based approach to combine the proposed two functionals and to mediate stabilization and safety objectives, which allows to achieve the stabilization objective under the safety requirement. The proposed approach is illustrated via a numerical example.
\end{abstract}

\begin{IEEEkeywords}
Control barrier functionals, Krasovskii approach, stochastic time-delay systems, safe stabilization.
\end{IEEEkeywords}

\section{Introduction}
\label{sec-intro}

In application domains like aviation, automobiles, energy and medicine \cite{Humayed2017cyber}, safety-critical systems involve some strict requirements on their states to avoid system damages and economic losses. Therefore, safety verification plays an important role in investigating whether dynamical systems work according to the specification requirements. For safety-critical systems, safety is in a priority place, whereas both constraint satisfaction and system stability are expected to be guaranteed simultaneously such that desired specifications can be accomplished. To ensure the system safety, many approaches have been proposed in the literature, such as model predictive control (MPC) \cite{Camacho2013model} and barrier function \cite{Ames2016control, Jankovic2018robust}. Different from MPC with inevitably heavy computational burden, barrier functions, similar to Lyapunov functions for the system stability \cite{Sontag1989universal, Pepe2017control, Scholl2022ode}, offer system-level certificates for the forward invariance of a specific region, which is called a ``safe set'' and can be associated with barrier functions. In this way, the freedom in the construction of barrier functions provides flexibility to deal with different safety constraints.

In existing works \cite{Ames2016control, Jankovic2018robust}, many types of barrier functions, including barrier Lyapunov functions \cite{Tee2009barrier}, zeroing/reciporal barrier functions based on the safe set \cite{Ames2016control} and barrier functions based on unsafe set \cite{Wieland2007constructive}, have been proposed to ensure the system safety from different perspectives. Currently, numerous efforts are made to deterministic or delay-free dynamical systems. However, due to information acquisition and computation for control decisions and executions \cite{Sipahi2011stability}, time delays and random noises are encountered inevitably, and may result in undesired issues like oscillation, instability and performance deterioration \cite{Gao2019stability}. Until now, based on barrier functions, the system safety has been studied for stochastic systems \cite{Santoyo2021barrier, Prajna2007framework, Clark2021control, Wisniewski2021safety} and time-delay systems \cite{Prajna2005methods, Orosz2019safety, Ren2022Razumikhin, Ren2022Razumikhin2, Jankovic2018control, Kiss2022control}. However, some limitations still exist in dealing with the safety problem under random noises and time delays. First, most barrier functions are constructed via the Razumikhin approach \cite{Jankovic2018control, Ren2022Razumikhin} and are only related to the current state. Only a few works consider the barrier functions in terms of the Krasovskii approach \cite{Orosz2019safety, Ren2022Razumikhin2, Kiss2022control}, where the time-delay trajectory is involved. Second, only the safety analysis is addressed for either the stochastic case \cite{Prajna2007framework, Clark2021control, Wisniewski2021safety} or the time-delay case \cite{Prajna2005methods, Orosz2019safety, Ren2022Razumikhin, Ren2022Razumikhin2}, while the safety control is difficult to be resolved via the quadratic programming \cite{Ames2016control, Jankovic2018robust} due to the computational complexity of time-delay optimization problems \cite{Wu2019new}. To the best of our knowledge, there exists few work on the safe stabilization problem in the stochastic time-delay case.

Motivated by the above discussion, in this paper we address the safe stabilization problem of stochastic nonlinear time-delay systems, and the Krasovskii approach is applied to involve the time-delay trajectory directly. To this end, we propose a stochastic control Lyapunov-Krasovskii functional (SCLKF) for the stabilization objective and a stochastic control barrier-Krasovskii functional (SCBKF) for the safety objective. In terms of the system stability, we further extend the small control property (SCP) \cite{Sontag1989universal} to the stochastic time-delay case. By combining the SCLKF and the stochastic SCP, a continuous stabilizing feedback controller is designed explicitly in a closed form. In terms of the system safety, the SCBKF is formulated and applied to guarantee the safety objective. Different from \cite{Orosz2019safety, Ren2022Razumikhin, Ren2022Razumikhin2, Kiss2022control} on the deterministic time-delay case, the SCLKF and SCBKF are established here for stochastic time-delay systems for the first time.

To achieve the stabilization and safety objectives simultaneously, we propose a sliding mode based approach to combine the SCLKF and SCBKF in a unified way. The motivation for this approach is that the optimization-based approach is not easy to be applied for stochastic time-delay systems due to the computational expense and the difficulties in finding a closed-form analytical solution. To deal with this issue, we establish a control framework based on the properties of stochastic sliding surface functionals, and derive a closed-form expression of the controller explicitly. Hence, the stabilization and safety objectives are guaranteed simultaneously for stochastic time-delay systems. In conclusion, comparing with many existing works \cite{Santoyo2021barrier, Prajna2007framework, Clark2021control, Wisniewski2021safety, Prajna2005methods, Orosz2019safety, Ren2022Razumikhin, Ren2022Razumikhin2, Jankovic2018control, Kiss2022control}, our main contributions are two-fold: (i) both SCLKF and SCBKF are proposed for the first time for stochastic time-delay systems; (ii) a sliding mode based approach is derived to combine the SCLKF and SCBKF and to design the controller to guarantee the stabilization and safety objectives simultaneously. Both contributions make a further step to extend the existing ones in \cite{Orosz2019safety, Ren2022Razumikhin, Ren2022Razumikhin2, Jankovic2018control, Kiss2022control} to the stochastic time-delay case, and are of great importance from a theoretical point of view as well as the potential practical application to the real-world dynamics with random noises.

Preliminaries are stated in Section \ref{sec-preliminary}. Stochastic control functionals are proposed in Section \ref{sec-functionals}. The controller is designed in Section \ref{sec-combination}. Numerical results are shown in Section \ref{sec-example}. Conclusions are given in Section \ref{sec-conclusion}.

\section{Preliminaries}
\label{sec-preliminary}

Let $\mathbb{R}:=(-\infty, +\infty), \mathbb{R}^{+}:=[0, +\infty), \mathbb{N}:=\{0, 1, \ldots\}$ and $\mathbb{N}^{+}:=\{1, 2, \ldots\}$. $|x|$ denotes the Euclidian norm of $x\in\mathbb{R}^{n}$, and $(x, y):=(x^{\top}, y^{\top})^{\top}$ for $x, y\in\mathbb{R}^{n}$. Given a set $\mathbb{A}\subset\mathbb{R}^{n}$, $\partial\mathbb{A}$ is its boundary, $\inte(\mathbb{A})$ is its interior, and $\overbar{\mathbb{A}}$ is its closure. An open ball centered at $\mathbf{x}\in\mathbb{R}^{n}$ with radius $\delta>0$ is denoted by $\mathbf{B}(\mathbf{x}, \delta):=\{x\in\mathbb{R}^{n}: |x-\mathbf{x}|<\delta\}$; $\mathbf{B}(\delta):=\mathbf{B}(0, \delta)$. $\mathcal{PC}([a, b], \mathbb{R}^{n})$ denotes the class of piecewise continuous functions mapping $[a, b]\subseteq\mathbb{R}$ to $\mathbb{R}^{n}$; $\mathcal{C}(\mathbb{R}^{n}, \mathbb{R}^{p})$ denotes the class of continuously differentiable functions mapping $\mathbb{R}^{n}$ to $\mathbb{R}^{p}$. Given $x\in\mathcal{PC}([-\Delta, +\infty), \mathbb{R}^{n})$, for any $t\in\mathbb{R}^{+}$, let $x_{t}$ be an element of $\mathcal{PC}([-\Delta, 0], \mathbb{R}^{n})$ defined by $x_{t}(\theta):=x(t+\theta)$ with $\theta\in[-\Delta, 0]$. For $\phi\in\mathcal{PC}([-\Delta, 0], \mathbb{R}^{n})$ with $\Delta>0$, we denote $\|\phi\|:=\sup_{\theta\in[-\Delta, 0]}|\phi(\theta)|$. For any $V\in\mathcal{C}(\mathbb{R}^{+}, \mathbb{R})$, its upper Dini derivative is $D^{+}V(t):=\limsup_{s\rightarrow0^{+}}\frac{V(t+s)-V(t)}{s}$. For any $h: \mathcal{PC}([-\Delta, 0], \mathbb{R}^{n})\rightarrow\mathbb{R}^{+}$, its upper Dini derivative is $D^{+}h(x_{t})=\limsup_{v\rightarrow0^{+}}\frac{h(x_{t+v})-h(x_{t})}{v}$. Let $\ep[\cdot]$ and $\tr[\cdot]$ denote the expectation and trace operators, respectively. A continuous function $\alpha: \mathbb{R}^{+}\rightarrow\mathbb{R}^{+}$ is of class $\mathcal{K}$ if it is strictly increasing and $\alpha(0)=0$; it is of class $\mathcal{K}_{\infty}$ if it is of class $\mathcal{K}$ and unbounded. A continuous function $\beta: \mathbb{R}^{+}\times\mathbb{R}^{+}\rightarrow\mathbb{R}^{+}$ is of class $\mathcal{KL}$ if, for each fixed $t\geq0$, $\beta(s, t)$ is of class $\mathcal{K}$, and for each fixed $s\geq0$, $\beta(s, t)$ decreases to 0 as $t\rightarrow\infty$.

\subsection{Stochastic Time-Delay Control Systems}
\label{subsec-systemodel}

In this paper, we consider stochastic nonlinear time-delay systems with the following dynamics:
\begin{align}
\label{eqn-1}
\begin{aligned}
dx(t)&=(f(x_{t})+g(x_{t})u)dt+\rho(x_{t})dw(t), \quad  t\geq0, \\
x(t)&=\xi(t), \quad t\in[-\Delta, 0],
\end{aligned}
\end{align}
where $x\in\mathbb{R}^{n}$ is the system state, $u\in\mathbb{U}\subset\mathbb{R}^{m}$ is the control input, and $w(t)\in\mathbb{R}^{p}$ is an $\mathfrak{F}_{t}$-adapted Brownian motion defined on a complete probability space $(\Omega, \mathfrak{F}, \pr, \{\mathfrak{F}_{t}\}_{t\geq0})$. Here we consider the case  $\mathbb{U}=\mathbb{R}^{m}$. That is, no constraint is imposed to the control input. $x_{t}\in\mathcal{PC}([-\Delta, 0], \mathbb{R}^{n})$ is the time-delay state, where $\Delta>0$ is the upper bound of time delays. The initial state is $\xi\in\mathcal{PC}([-\Delta, 0], \mathbb{X}_{0})$, where $\mathbb{X}_{0}\subset\mathbb{R}^{n}$ includes the origin. $\ep[\|\xi\|]$ is assumed to be bounded, where $\|\xi\|:=\sup_{\theta\in[-\Delta, 0]}|\xi(\theta)|$. Assume that the functionals $f: \mathcal{PC}([-\Delta, 0], \mathbb{R}^{n})\rightarrow\mathbb{R}^{n}, g: \mathcal{PC}([-\Delta, 0], \mathbb{R}^{n})\rightarrow\mathbb{R}^{n\times m}$ and $\rho: \mathcal{PC}([-\Delta, 0], \mathbb{R}^{n})\rightarrow\mathbb{R}^{n\times p}$ are continuous and locally Lipschitz, which ensures the existence of the unique solution to the system \eqref{eqn-1}; see \cite[Sec. 5]{Mao2007stochastic} for more details. Let $f(0)=0, g(0)=0$ and $\rho(0)=0$. Hence, $x(t)\equiv0$ for all $t>0$ is a trivial solution to the system \eqref{eqn-1}.

\begin{definition}[\cite{Ren2016stability}]
\label{def-1}
Given the input $u\in\mathbb{U}$, the system \eqref{eqn-1} is \emph{stochastically globally asymptotically stable (SGAS)}, if for any $\varepsilon\in(0, 1)$, there exists $\beta\in\mathcal{KL}$ such that $\pr\{|x(t)|\leq\beta(\ep[\|\xi\|], t)\}\geq1-\varepsilon$ for all $t\geq0$ and $\xi\in\mathcal{PC}([-\Delta, 0], \mathbb{X}_{0})$.
\end{definition}

Based on Definition \ref{def-1}, the \textit{stabilization control} is to design a controller such that the system \eqref{eqn-1} is SGAS. On the other hand, the \textit{safety control} is to design a controller such that the system \eqref{eqn-1} stays in a predefined safe set $\mathbb{S}\subset\mathcal{PC}([-\Delta, 0], \mathbb{R}^{n})$. The safe set $\mathbb{S}$ is associated with a continuously differentiable functional $h: \mathcal{PC}([-\Delta, 0], \mathbb{R}^{n})\rightarrow\mathbb{R}$. That is,
\begin{align}
\label{eqn-2}
\mathbb{S}&:=\{\phi\in\mathcal{PC}([-\Delta, 0], \mathbb{R}^{n}): h(\phi)\geq0\}, \\
\label{eqn-3}
\partial\mathbb{S}&:=\{\phi\in\mathcal{PC}([-\Delta, 0], \mathbb{R}^{n}): h(\phi)=0\}, \\
\label{eqn-4}
\inte(\mathbb{S})&:=\{\phi\in\mathcal{PC}([-\Delta, 0], \mathbb{R}^{n}): h(\phi)>0\}.
\end{align}
Let $\inte(\mathbb{S})\neq\varnothing$ and $\overbar{\inte(\mathbb{S})}=\mathbb{S}$. To address the stabilization and safety objectives, the following notation is defined.

\begin{definition}[\cite{Mao2007stochastic}]
\label{def-2}
Given any continuously differentiable function $V: \mathbb{R}^{n}\rightarrow\mathbb{R}^{+}$, \emph{the infinitesimal operator} of $V(x)$, associated with the system \eqref{eqn-1}, is defined as
\begin{equation*}
\mathcal{L}V(\phi):=L_{f}V(\phi)+L_{g}V(\phi)u+\frac{1}{2}\tr\left[\rho^{\top}(\phi)\frac{\partial^{2}V(x)}{\partial x^{2}}\rho(\phi)\right],
\end{equation*}
where $\phi\in\mathcal{PC}([-\Delta, 0], \mathbb{R}^{n})$, $L_{f}V(\phi):=\frac{\partial V(x)}{\partial x}f(\phi)$ and $L_{g}V(\phi):=\frac{\partial V(x)}{\partial x}g(\phi)$.
\end{definition}

In order to facilitate the following analysis, from Definition \ref{def-2}, we introduce the following notation:
\begin{equation*}
\mathcal{L}_{a}V(\phi):=L_{f}V(\phi)+0.5\tr\left[\rho^{\top}(\phi)\frac{\partial^{2}V(x)}{\partial x^{2}}\rho(\phi)\right].
\end{equation*}

\section{Stochastic Control Functionals}
\label{sec-functionals}

In this section we propose stochastic control functionals for the system \eqref{eqn-1}. We present the stochastic control Lyapunov-Krasovskii functional in Section \ref{subsec-SCLKF}, and then the stochastic control barrier-Krasovskii functional in Section \ref{subsec-SCBKF}.

\subsection{Stochastic Control Lyapunov-Krasovskii Functional}
\label{subsec-SCLKF}

To establish stochastic control Lyapunov-Krasovskii functionals for \eqref{eqn-1}, we recall smoothly separable functionals.

\begin{definition}[\cite{Pepe2014stabilization}]
\label{def-3}
A functional $V: \mathcal{PC}([-\Delta, 0], \mathbb{R}^{n})\rightarrow\mathbb{R}^{+}$ is \emph{smoothly separable}, if there exist $V_{1}\in\mathcal{C}(\mathbb{R}^{n}, \mathbb{R}^{+})$, a locally Lipschitz functional $V_{2}: \mathcal{PC}([-\Delta, 0], \mathbb{R}^{n})\rightarrow\mathbb{R}^{+}$, and $\alpha_{1}, \alpha_{2}\in\mathcal{K}_{\infty}$ such that, for all $\phi\in\mathcal{PC}([-\Delta, 0], \mathbb{R}^{n})$,
\begin{align}
\label{eqn-5}
V(\phi)&:=V_{1}(\phi(0))+V_{2}(\phi), \\
\label{eqn-6}
\alpha_{1}(|\phi(0)|)&\leq V_{1}(\phi(0))\leq\alpha_{2}(|\phi(0)|).
\end{align}
\end{definition}

From Definition \ref{def-3}, any smoothly separable functional is locally Lipschitz. For any smoothly separable functional, we introduce its property of stochastic invariant differentiability, which will be used to the controller design afterwards.

\begin{definition}[\cite{Kim1999functional}]
\label{def-4}
A smoothly separable functional $V:\mathcal{PC}([-\Delta, 0], \mathbb{R}^{n})\rightarrow\mathbb{R}^{+}$ is \emph{invariantly differentiable (i-differentiable)}, if $V(\phi)=V_{1}(\phi(0))+V_{2}(\phi)$ and
\begin{enumerate}[(1)]
  \item for any $\phi\in\mathcal{PC}([-\Delta, 0], \mathbb{R}^{n})$ with $x=\phi(0)$, both $\partial V_{1}(x)/\partial x$ and $D^{+}V_{2}(\phi)$ exist;
  \item $D^{+}V_{2}(\phi)$ is invariant with respect to $\phi\in\mathcal{PC}([-\Delta, 0], \mathbb{R}^{n})$, that is, $D^{+}V_{2}(x_{0})$ is the same for all $x_{t}\in\mathcal{PC}([-\Delta, 0], \mathbb{R}^{n})$;
  \item for all $x_{t}\in\mathcal{PC}([-\Delta, 0], \mathbb{R}^{n})$ and $l\geq0$, $V(x_{t+l})-V(x_{t}):=\frac{\partial V_{1}(y)}{\partial y}z+D^{+}V_{2}(x_{t})l+o(\sqrt{|z|^{2}+l^{2}})$, where $y=x_{t}(0), z=x_{t+l}(0)-x_{t}(0)$ and $\lim_{s\rightarrow0^{+}}o(s)/s=0$.
\end{enumerate}
In addition, if $D^{+}V_{2}(\phi)$ is continuous, then $V$ is said to be \emph{continuously i-differentiable}.
\end{definition}

In Definitions \ref{def-3}-\ref{def-4}, $V_{2}$ is defined on $\mathcal{PC}([-\Delta, 0], \mathbb{R}^{n})$ to ensure the well-posedness of $D^{+}V_{2}(\phi)$, which is different from \cite{Kim1999functional} where the invariant differentiability is for the functionals on $\mathbb{R}^{n}\times\mathcal{PC}([-\Delta, 0), \mathbb{R}^{n})$. Definitions \ref{def-3}-\ref{def-4} include many types of Lyapunov-Krasovskii functionals \cite{Kim1999functional}. From Definition \ref{def-4} and the It\^{o}'s differential formula in \cite[Ch. 1]{Mao2007stochastic},
\begin{align*}
dV(\phi)&=(\mathcal{L}V_{1}(\phi)+D^{+}V_{2}(\phi))dt+\frac{\partial V_{1}(\phi(0))}{\partial\phi(0)}\rho(\phi)dw(t). 
\end{align*}
From the property of the random noise, we have $d\ep[V(\phi)]=\ep[\mathcal{L}V_{1}(\phi)+D^{+}V_{2}(\phi)]dt$. Moreover, from \cite{Zhao2012stochastic}, if $\ep[\mathcal{L}V_{1}(\phi)+D^{+}V_{2}(\phi)]$ is continuous, then we can define $\mathcal{V}(t):=\ep[V(x_{t})]$ and obtain $D^{+}\mathcal{V}(t)=\ep[\mathcal{L}V_{1}(\phi)+D^{+}V_{2}(\phi)]$. We next propose the stochastic control Lyapunov-Krasovskii functional and the stochastic small control property.

\begin{definition}
\label{def-5}
For the system \eqref{eqn-1}, a continuously i-differentia-ble functional $V: \mathcal{PC}([-\Delta, 0], \mathbb{R}^{n})\rightarrow\mathbb{R}^{+}$ is called a \emph{stocha-stic control Lyapunov-Krasovskii functional (SCLKF)}, if
\begin{enumerate}[(i)]
  \item there exist $\alpha_{1}, \alpha_{2}\in\mathcal{K}_{\infty}$ such that $\alpha_{1}(|\phi(0)|)\leq V(\phi)\leq\alpha_{2}(\|\phi\|)$ for all $\phi\in\mathcal{PC}([-\Delta, 0], \mathbb{R}^{n})$,

  \item there exists $\gamma_{1}\in\mathcal{K}$ such that $\inf_{u\in\mathbb{U}}\{\mathcal{L}V_{1}(\phi)+D^{+}V_{2}(\phi)\}<-\gamma_{1}(V(\phi))$ for any nonzero $\phi\in\mathcal{PC}([-\Delta, 0], \mathbb{R}^{n})$. 
\end{enumerate}
\end{definition}

\begin{definition}
\label{def-6}
Consider the system \eqref{eqn-1} with an SCLKF $V:\mathcal{PC}([-\Delta, 0], \mathbb{R}^{n})\rightarrow\mathbb{R}^{+}$. The system \eqref{eqn-1} is said to satisfy the \emph{stochastic small control property (SSCP)}, if for arbitrary $\varepsilon>0$, there exists $\delta>0$ such that, for any nonzero $\phi\in\mathcal{PC}([-\Delta, 0], \mathbf{B}(\delta))$, there exists $u\in\mathbf{B}(\varepsilon)$ such that $\mathcal{L}V_{1}(\phi)+D^{+}V_{2}(\phi)<-\gamma_{1}(V(\phi))$.
\end{definition}

Definition \ref{def-5} provides a novel control Lyapunov functional for the system \eqref{eqn-1}, and extends these in \cite{Pepe2014stabilization, Ren2022Razumikhin} to the stochastic case. Similarly, Definition \ref{def-6} extends the small control property to the stochastic time-delay case via the SCLKF in Definition \ref{def-5}. With the proposed SCLKF and SSCP, the feedback controller is derived in the following theorem to ensure the SGAS property of the system \eqref{eqn-1}.

\begin{theorem}
\label{thm-1}
If the system \eqref{eqn-1} admits an SCLKF and satisfies the SSCP, then the following continuous controller
\begin{align}
\label{eqn-7}
u(\phi):=\left\{\begin{aligned}
&\kappa(\lambda, \mathfrak{a}(\phi), (L_{g}V_{1}(\phi))^{\top}), &\text{ if }& \phi\neq 0, \\
&0, &\text{ if }&  \phi=0,
\end{aligned}\right.
\end{align}
ensures the SGAS property of the system \eqref{eqn-1}, where $\lambda>0$, $\mathfrak{a}(\phi):=\mathcal{L}_{a}V_{1}(\phi)+D^{+}V_{2}(\phi)+\gamma_{1}(V(\phi))$, and
\begin{equation}
\label{eqn-8}
\kappa(\lambda, p, q)=\left\{\begin{aligned}
&\frac{\mathfrak{p}+\sqrt{\mathfrak{p}^{2}+\lambda\|\mathfrak{q}\|^{4}}}{-\|\mathfrak{q}\|^{2}}\mathfrak{q}, &\text{ if }\ & \mathfrak{q}\neq 0, \\
&0, &\text{ if }\ &  \mathfrak{q}=0.
\end{aligned}\right.
\end{equation}
\end{theorem}

\begin{IEEEproof}
We first show the stabilization of the system \eqref{eqn-1} under the controller \eqref{eqn-7}. From \eqref{eqn-7}, if $L_{g}V_{1}(\phi)=0$, then $u=0$, and from Definition \ref{def-5},
\begin{align*}
&\mathcal{L}V_{1}(\phi)+D^{+}V_{2}(\phi)+\gamma_{1}(V(\phi)) \\
&=\mathcal{L}_{a}V_{1}(\phi)+D^{+}V_{2}(\phi)+\gamma_{1}(V(\phi))<0,
\end{align*}
where $\mathcal{L}V_{1}(\phi)$ and $\mathcal{L}_{a}V_{1}(\phi)$ are defined in Section \ref{subsec-systemodel}. If $L_{g}V_{1}(\phi)\neq0$, then let $\mathfrak{b}(\phi):=L_{g}V_{1}(\phi)$, and from \eqref{eqn-7},
\begin{align*}
&\mathcal{L}V_{1}(\phi)+D^{+}V_{2}(\phi)+\gamma_{1}(V(\phi)) \\
&=\mathfrak{a}(\phi)-\mathfrak{b}(\phi)\frac{\mathfrak{a}(\phi)\mathfrak{b}^{\top}(\phi)}{\|\mathfrak{b}(\phi)\|^{2}} -\mathfrak{b}(\phi)\frac{\sqrt{\mathfrak{a}^{2}(\phi)+\lambda\|\mathfrak{b}(\phi)\|^{4}}}{\|\mathfrak{b}(\phi)\|^{2}}\mathfrak{b}^{\top}(\phi) \\
&=-\sqrt{\mathfrak{a}^{2}(\phi)+\lambda\|\mathfrak{b}(\phi)\|^{4}}\leq0.
\end{align*}
Hence, under the controller \eqref{eqn-7}, we have that for all $t\geq0$,
\begin{align}
\label{eqn-9}
\mathcal{L}V_{1}(\phi)+D^{+}V_{2}(\phi)&<-\gamma_{1}(V(\phi))\leq-\gamma_{1}(\alpha_{1}(\phi(0))),
\end{align}
where the second ``$\leq$'' holds from item (i) in Definition \ref{def-5}. From \eqref{eqn-9} and \cite[Thm. 2]{Liu2008adaptive}, there exists $\beta_{1}\in\mathcal{KL}$ such that
\begin{align}
\label{eqn-10}
\ep[V(\phi)]&\leq\beta_{1}(\alpha_{2}(\ep[\|\xi\|]), t), \quad \forall t\geq0.
\end{align}
Using Markov's inequality in \cite[Ch. II, 18.1]{Rogers2000diffusions1}, we derive from \eqref{eqn-10} and  item (i) in Definition \ref{def-5} that
\begin{align*}
\pr\left\{|x(t)|\leq\beta(\ep[\|\xi\|], t)\right\}\geq1-\varepsilon, \quad  \forall \varepsilon\in(0, 1), \forall t\geq0,
\end{align*}
where $\beta(v, t):=\alpha^{-1}_{1}(\varepsilon^{-1}\beta_{1}(\alpha_{2}(v), t))\in\mathcal{KL}$. Therefore, we conclude that the system \eqref{eqn-1} is SGAS.

Next, we show the continuity of the controller \eqref{eqn-7}. From the properties of the functional $V$ and the functionals $f, g$, we can deduce the continuity of the controller in any region away from the origin. Hence, we only need to show the continuity of the controller at the origin. From the SSCP, for arbitrary $\varepsilon\in\mathbb{R}^{+}$, there exists $\delta_{1}\in\mathbb{R}^{+}$ such that, for any nonzero $\phi\in\mathcal{PC}([-\Delta, 0], \mathbf{B}(\delta_{1}))$, there exists $u\in\mathbf{B}(\varepsilon)$ such that $\mathfrak{a}(\phi)+\mathfrak{b}(\phi)u<0$. Since $V_{1}\in\mathcal{C}(\mathbb{R}^{n}, \mathbb{R}^{+})$ and $g$ in \eqref{eqn-1} is locally Lipschitz, there exists $\delta_{2}\in\mathbb{R}^{+}$ with $\delta_{2}\neq\delta_{1}$ such that $\|\mathfrak{b}(\phi)\|\leq\varepsilon$ holds for all nonzero $\phi\in\mathcal{PC}([-\Delta, 0], \mathbf{B}(\delta_{2}))$. Let $\delta:=\min\{\delta_{1}, \delta_{2}\}$, and for any nonzero $\phi\in\mathcal{PC}([-\Delta, 0], \mathbf{B}(\delta))$, $\|\mathfrak{b}(\phi)\|\leq\varepsilon$ and there exists $u\in\mathbf{B}(\varepsilon)$ such that $\mathfrak{a}(\phi)+\mathfrak{b}(\phi)u<0$.

For any nonzero $\phi\in\mathcal{PC}([-\Delta, 0], \mathbf{B}(\delta))$, we consider the following two cases. The first case is $\mathfrak{b}(\phi)=0$. In this case, $u(\phi)=0$ from \eqref{eqn-8}. In addition, $u(0)=0$ from \eqref{eqn-7}, and thus $\|u(\phi)-u(0)\|=0<\varepsilon$. Hence, if the SSCP is satisfied, then the control input is bounded by $\varepsilon\in\mathbb{R}^{+}$. Since $\varepsilon\in\mathbb{R}^{+}$ can be arbitrarily small, the controller \eqref{eqn-7} is continuous in this case. The second case is $\mathfrak{b}(\phi)\neq0$. In this case, $|\mathfrak{a}(\phi)|\leq\varepsilon\|\mathfrak{b}(\phi)\|$ for any nonzero $\phi\in\mathcal{PC}([-\Delta, 0], \mathbf{B}(\delta))$. From \eqref{eqn-7}, we have for any nonzero $\phi\in\mathcal{PC}([-\Delta, 0], \mathbf{B}(\delta))$,
\begin{align*}
\|u(\phi)\|&\leq\left\|\frac{\mathfrak{a}(\phi)+\sqrt{\mathfrak{a}^{2}(\phi)+\lambda\|\mathfrak{b}(\phi)\|^{4}}}{\|\mathfrak{b}(\phi)\|^{2}}\mathfrak{b}(\phi)\right\| \\
&\leq\left|\frac{\mathfrak{a}(\phi)+\sqrt{\mathfrak{a}^{2}(\phi)+\lambda\|\mathfrak{b}(\phi)\|^{4}}}{\|\mathfrak{b}(\phi)\|}\right|  \\
&\leq\left|\frac{\mathfrak{a}(\phi)+|\mathfrak{a}(\phi)|+\sqrt{\lambda}\|\mathfrak{b}(\phi)\|^{2}}{\|\mathfrak{b}(\phi)\|}\right|\leq(2+\sqrt{\lambda})\varepsilon.
\end{align*}
Since $\lim_{\varepsilon\rightarrow0}(2+\sqrt{\lambda})\varepsilon=0$ and $\varepsilon\in\mathbb{R}^{+}$ can be arbitrarily small, the controller \eqref{eqn-7} is continuous at the origin in the second case. As a result, we conclude that the controller \eqref{eqn-7} is continuous at the origin and the proof is completed.
\end{IEEEproof}

Theorem \ref{thm-1} extends the classic Sontag's formula \cite{Sontag1989universal} into the stochastic time-delay case. We stress that the controller \eqref{eqn-7} is continuous for the case of $\phi\in\mathcal{PC}([-\Delta, 0], \mathbb{R}^{n}\setminus\{0\})$ and the case of $\phi\in\mathcal{PC}([-\Delta, 0], \{0\})$. Furthermore, the controller \eqref{eqn-7} is general enough and the functional $g$ in \eqref{eqn-1} is not necessarily required to be zero at the origin.

\subsection{Stochastic Control Barrier-Krasovskii Functional}
\label{subsec-SCBKF}

To investigate the safety of the system \eqref{eqn-1}, the stochastic control barrier-Krasovskii functional is proposed in this subsection. For this purpose, we start with the property of the stochastic forward invariance.

\begin{definition}
\label{def-7}
A compact set $\mathbb{A}\subset\mathbb{R}^{n}$ is said to be \emph{stochastically forward invariant} for the system \eqref{eqn-1}, if for arbitrary $\varepsilon\in(0, 1)$ and all $\xi\in\mathcal{PC}([-\Delta, 0], \mathbb{A})$, $\pr\{x(t)\in\mathbb{A}\}\geq1-\varepsilon$ holds for all $t\geq0$.
\end{definition}

From Definition \ref{def-7}, for a given probability $1-\varepsilon$, there exists an initial condition such that the resulting state trajectory stays in $\mathbb{A}\subset\mathbb{R}^{n}$ at least with the probability $1-\varepsilon$. From Definition \ref{def-7}, the set $\mathbb{S}$ in \eqref{eqn-2} is stochastically forward invariant for the system \eqref{eqn-1}, if $\pr\{x_{t}\in\mathbb{S}\}\geq1-\varepsilon$ holds for arbitrary $\varepsilon\in(0, 1)$ and all $t\geq0$, or equivalently $\pr\{h(x_{t})\geq0\}\geq1-\varepsilon$ holds for arbitrary $\varepsilon\in(0, 1)$ and all $t\geq0$. Next, the stochastic control barrier-Krasovskii functional is presented.

\begin{definition}
\label{def-8}
Consider the system \eqref{eqn-1} and the set $\mathbb{S}$ in \eqref{eqn-2}. A continuously i-differentiable functional $B: \inte(\mathbb{S})\rightarrow\mathbb{R}$ is called an \emph{stochastic control barrier-Krasovskii functional (SCBKF)} for the set $\mathbb{S}$, if
\begin{enumerate}[(i)]
\item there exist $\alpha_{1}, \alpha_{2}\in\mathcal{K}$ such that for all $\phi\in\inte(\mathbb{S})$, $\alpha_{1}(h(\phi))\leq1/B(\phi)\leq\alpha_{2}(h(\phi))$;

\item there exists $\gamma_{2}\in\mathcal{K}$ such that $\inf_{u\in\mathbb{U}}\{\mathcal{L}B_{1}(\phi)+D^{+}B_{2}(\phi)\}<\gamma_{2}(h(\phi))$ holds for all nonzero $\phi\in\inte(\mathbb{S})$.
\end{enumerate}
\end{definition}

From Definition \ref{def-8}, we define the following set
\begin{align}
\label{eqn-11}
\mathbb{K}&:=\{u\in\mathbb{U}: \mathcal{L}B_{1}(\phi)+D^{+}B_{2}(\phi)<\gamma_{2}(h(\phi))\}.
\end{align}
We next show that the control inputs from the set $\mathbb{K}$ result in the stochastically forward invariant property of the set $\mathbb{S}$.

\begin{theorem}
\label{thm-2}
Consider the system \eqref{eqn-1} and the set $\mathbb{S}$ in \eqref{eqn-2}. If the system \eqref{eqn-1} admits an SCBKF $B: \inte(\mathbb{S})\rightarrow\mathbb{R}$, and there exists a Lipschitz continuous functional $u: \inte(\mathbb{S})\rightarrow\mathbb{U}$ such that $u\in\mathbb{K}$ with $\mathbb{K}$ in \eqref{eqn-11}, then $u$ is a controller to ensure the stochastically forward invariant property of the set $\inte(\mathbb{S})$.
\end{theorem}

\begin{IEEEproof}
For the SCBKF $B$, we define an functional $\Theta(x_{t}):=1/B(x_{t})$. Hence, $D^{+}\Theta(\phi)=-D^{+}B(\phi)/(B^{2}(\phi))$ and from Jensen's inequality,  $1/\ep[B(\phi)]\leq\ep[\Theta(\phi)]\leq\ep[\alpha_{2}(h(\phi))]$. Let $\mathcal{B}(t):=\ep[\Theta(x_{t})]$, and
\begin{align}
\label{eqn-12}
D^{+}\mathcal{B}(t)&=-\ep[\Theta^{2}(\phi)(\mathcal{L}B_{1}(\phi)+D^{+}B_{2}(\phi))] \nonumber\\
&>-\ep[\Theta^{2}(\phi)\gamma_{2}(h(\phi))] \nonumber \\
&\geq-\ep[\Theta^{2}(\phi)\gamma_{2}(\alpha^{-1}_{1}(\Theta(\phi)))].
\end{align}
Let $\varphi(\Theta(\phi)):=\Theta^{2}(\phi)\gamma_{2}(\alpha^{-1}_{1}(\Theta(\phi)))$, which is of class $\mathcal{K}$. From \cite[Thm. 3.1]{Pepe2006lyapunov} and the comparison principle, there exists $\zeta\in\mathcal{KL}$ such that
\begin{align}
\label{eqn-13}
\mathcal{B}(t)\geq\zeta(\mathcal{B}(0), t), \quad  \forall t\geq0.
\end{align}
Combining \eqref{eqn-13} with the definition of $\Theta(\phi)$ implies
\begin{align*}
\ep[1/B(x_{t})]&\geq\zeta(\ep[1/B(x_{0})], t), \quad  \forall t\geq0,
\end{align*}
which further implies from item (i) of Definition \ref{def-8} that
\begin{equation}
\label{eqn-14}
\ep[\alpha_{2}(h(x_{t}))]\geq\zeta(\alpha_{1}(h(\ep[\|\xi\|])), t), \quad  \forall t\geq0.
\end{equation}
From \eqref{eqn-14}, it is easy to check that $\ep[\alpha_{2}(h(x_{t}))]>0$ for all $t\geq0$. Let $\beta(v, t):=\zeta(\alpha_{1}(h(v)), t)$, which is of class $\mathcal{KL}$. From Chebyshev's inequality \cite[Ch. 3]{Motwani1995randomized} and \eqref{eqn-14}, we obtain that for arbitrary $\varepsilon\in(0, 1)$ and all $t\geq0$,
\begin{align}
\label{eqn-15}
&\pr\left\{\alpha_{2}(h(x_{t}))\geq\ep[\alpha_{2}(h(x_{t}))]-\sqrt{\varepsilon^{-1}\ep[\alpha_{2}(h(x_{t}))]}\right\} \nonumber  \\
&\geq1-\varepsilon.
\end{align}
Since $\ep[\alpha_{2}(h(x_{t}))]>0$ holds for all $t\geq0$, from \eqref{eqn-15} we consider the following two cases: $\varepsilon\ep[\alpha_{2}(h(x_{t}))]>1$ and $\varepsilon\ep[\alpha_{2}(h(x_{t}))]\leq1$. In the first case, we have from \eqref{eqn-15} that
\begin{align*}
\pr\{\alpha_{2}(h(x_{t}))>0\}\geq1-\varepsilon,
\end{align*}
which further shows that $\pr\{h(x_{t})>0\}\geq1-\varepsilon$. In the second case, $\ep[\alpha_{2}(h(x_{t}))]\leq\varepsilon^{-1}$, combining which with the Chernoff bound \cite[Ch. 4]{Motwani1995randomized} yields that, for all $\delta\in(0, 1)$,
\begin{align*}
\pr\{\alpha_{2}(h(x_{t}))\leq(1-\delta)\varepsilon^{-1}\}\geq e^{-0.5\delta^{2}\ep[h(x_{t})]},
\end{align*}
which further implies
\begin{align*}
\pr\{\alpha_{2}(h(x_{t}))\geq(1-\delta)\varepsilon^{-1}\}\geq1-e^{-0.5\delta^{2}\varepsilon^{-1}}.
\end{align*}
Let $\delta=1-\varepsilon^{2}\in(0, 1)$, and then $\pr\{\alpha_{2}(h(x_{t}))\geq\varepsilon\}\geq1-\exp(-0.5(1-\varepsilon^{2})^{2}\varepsilon^{-1})$, which further indicates that
$\pr\{h(x_{t})\geq\alpha^{-1}_{2}(\varepsilon)\}\geq1-\exp(-0.5(1-\varepsilon^{2})^{2}\varepsilon^{-1})$. We denote $\mu:=\max\{\varepsilon, \exp(-0.5(1-\varepsilon^{2})^{2}\varepsilon^{-1})\}\in(0, 1)$, and have $\pr\{h(x_{t})>0\}\geq1-\mu$ for all $t\geq0$, which implies that the set $\inte(\mathbb{S})$ is stochastically forward invariant.
\end{IEEEproof}

From Theorem \ref{thm-2}, the existence of the SCBKF implies the stochastic forward invariance of the set $\mathbb{S}$. Different from \cite{Clark2019control} on stochastic systems and \cite{Ren2022Razumikhin} on time-delay systems, Theorem \ref{thm-2} offers a novel result for the stochastic time-delay case. Comparing with \cite{Clark2019control} using the techniques of martingales and stopping times, a simple and direct proof is presented here based on the comparison principle and tail inequalities \cite{Motwani1995randomized}.

\section{Sliding Mode-based Controller Design}
\label{sec-combination}

To mediate the safety and stabilization objectives of the system \eqref{eqn-1}, the SCLKF and SCLBF are combined together in this section. Since time-delay optimal control problems are not easy to be resolved, the combination is based on the sliding-mode approach. Let the SCLKF be $V$ and the SCBKF be $B$. The sliding surface functional is defined below.
\begin{align}
\label{eqn-16}
U(\phi)&:=\psi(V(\phi), B(\phi)),
\end{align}
where the functional $U: \mathcal{PC}([-\Delta, 0], \mathbb{R}^{n})\rightarrow\mathbb{R}$ and the function $\psi: \mathbb{R}\times\mathbb{R}\rightarrow\mathbb{R}$ are assumed to be continuously differentiable. Define $\mathcal{U}(t):=\ep[U(x_{t})]$, and thus
\begin{equation}
\label{eqn-17}
D^{+}\mathcal{U}(t):=\mathbf{F}(\phi)+\mathbf{G}(\phi)u+\mathbf{L}(\phi)
\end{equation}
with $\mathbf{F}(\phi)=\mathbf{H}(\phi)f(\phi), \mathbf{G}(\phi)=\mathbf{H}(\phi)g(\phi)$ and
\begin{align*}
\mathbf{H}(\phi)&=\frac{\partial\psi}{\partial{V}}\frac{\partial{V}_{1}(\phi(0))}{\partial\phi(0)}+\frac{\partial\psi}{\partial{B}}\frac{\partial{B}_{1}(\phi(0))}{\partial\phi(0)}, \\ \mathbf{L}(\phi)&=\frac{\partial\psi}{\partial{V}}D^{+}{V}_{2}(\phi)+\frac{\partial\psi}{\partial{B}}D^{+}{B}_{2}(\phi) \\
&\quad +\frac{1}{2}\tr\left[\rho^{\top}(\phi)\left(\frac{\partial^{2}V_{1}(\phi(0))}{\partial\phi^{2}(0)}+\frac{\partial^{2}B_{1}(\phi(0))}{\partial \phi^{2}(0)}\right)\rho(\phi)\right].
\end{align*}
The following assumption, which is called \emph{the transversality condition} \cite{Sira1999general}, is to avoid $g(\phi)$ to be orthogonal to $\mathbf{H}(\phi)$.

\begin{assumption}
\label{asp-1}
For all $\phi\in\mathcal{PC}([-\Delta, 0], \mathbb{R}^{n})$, $\mathbf{G}(\phi)\neq0$.
\end{assumption}

From Assumption \ref{asp-1}, $g(\phi)$ is not tangential to the level set of the sliding surface functional $U(\phi)$. If Assumption \ref{asp-1} does not hold, then higher-order sliding surface functionals can be introduced \cite{Levant1993sliding} to guarantee that the following analysis can be proceeded similarly. Based on \eqref{eqn-17}, the following auxiliary functionals are introduced.
\begin{align*}
\mathbf{J}_{1}(\phi)&=\frac{g(\phi)\mathbf{G}^{\top}(\phi)f^{\top}(\phi)-f(\phi)\mathbf{G}(\phi)g^{\top}(\phi)}{2\|\mathbf{G}(\phi)\|^{2}}, \\
\mathbf{J}_{2}(\phi)&=\frac{g(\phi)\mathbf{G}^{\top}(\phi)f^{\top}(\phi)+f(\phi)\mathbf{G}(\phi)g^{\top}(\phi)}{2\|\mathbf{G}(\phi)\|^{2}}.
\end{align*}
Since $\mathbf{H}(\phi)g(\phi)g^{\top}(\phi)\mathbf{H}^{\top}(\phi)$ is symmetric and $\mathbf{H}(\phi)f(\phi)\in\mathbb{R}$, we can check that $\mathbf{H}(\phi)(\mathbf{J}_{1}(\phi)+\mathbf{J}_{2}(\phi))\mathbf{H}^{\top}(\phi)=\mathbf{F}(\phi)$ and $\mathbf{H}(\phi)\mathbf{J}_{1}(\phi)\mathbf{H}^{\top}(\phi)=0$. 

With these preliminaries, we next address the controller design. In the ideal sliding motion case, the system trajectory is to satisfy the manifold invariant condition $U(\phi)=0$, which can be verified via the functional $0.5U^{2}(\phi)$. Let $\mathcal{W}(t):=0.5\ep[U^{2}(x_{t})]$ and then
\begin{align}
\label{eqn-18}
&D^{+}\mathcal{W}(t)=\ep[U(\phi)D^{+}U(\phi)]\nonumber\\
&\quad  =\ep[U(\phi)(-\mathbf{H}(\phi)\mathbf{J}_{1}(\phi)\mathbf{H}^{\top}(\phi)+\mathbf{H}(\phi)\mathbf{J}_{2}(\phi)\mathbf{H}^{\top}(\phi) \nonumber\\
&\qquad  +\mathbf{L}(\phi)+\mathbf{G}(\phi)u)] \nonumber\\
&\quad  =\ep[U(\phi)(\mathbf{H}(\phi)\mathbf{J}_{2}(\phi)\mathbf{H}^{\top}(\phi)+\mathbf{L}(\phi)+\mathbf{G}(\phi)u)].
\end{align}
From $D^{+}\mathcal{W}(t)=0$, the ideal controller is derived as
\begin{align*}
u_{\mathsf{e}}(\phi)=\frac{\mathbf{G}^{\top}(\phi)(\mathbf{H}(\phi)\mathbf{J}_{2}(\phi)\mathbf{H}^{\top}(\phi)+\mathbf{L}(\phi))}{-\|\mathbf{G}(\phi)\|^{2}}.
\end{align*}
Since the exact system state may move into the sublevel and superlevel sets of the sliding surface, the applied controller is 
\begin{align}
\label{eqn-19}
u(\phi)=\frac{\mathbf{G}^{\top}(\phi)(\mathbf{H}(\phi)\mathbf{J}_{2}(\phi)\mathbf{H}^{\top}(\phi)+\mathbf{L}(\phi)+\mathbf{K}(\phi))}{-\|\mathbf{G}(\phi)\|^{2}},
\end{align}
where $\mathbf{K}(\phi)>0$ is an additional item to be designed. Note that $u(\phi)=u_{\mathsf{e}}(\phi)-\mathbf{G}^{\top}(\phi)\mathbf{K}(\phi)/\|\mathbf{G}(\phi)\|^{2}$. With the controller \eqref{eqn-19}, the following theorem is derived to guarantee the safe stabilization of the system \eqref{eqn-1}.

\begin{theorem}
\label{thm-3}
Consider the system \eqref{eqn-1} with the safe set $\mathbb{S}\subset\mathcal{PC}([-\Delta, 0], \mathbb{R}^{n})$ defined in \eqref{eqn-2}-\eqref{eqn-4}. Let $\xi\in\inte(\mathbb{S})$. If Assumption \ref{asp-1} holds, and the functional $U$ in \eqref{eqn-16} is such that
\begin{align}
\label{eqn-20}
\ep[U^{2}(\phi)]\geq\ep[U^{2}(\xi)], \quad \forall \phi\in\partial\mathbb{S}, \\
\label{eqn-21}
\mathbb{A}:=\{\phi\in\mathbb{S}: U(\phi)=0\}\subset\inte(\mathbb{S}),
\end{align}
then the stabilization and safety objectives can be achieved simultaneously via the controller \eqref{eqn-19} with
\begin{equation}
\label{eqn-22}
\mathbf{K}(\phi):=\frac{\mathsf{K}U(\phi)}{\|U(\phi)\|+\varpi},
\end{equation}
where $\mathsf{K}>0$ is constant and $\varpi>0$ can be sufficiently small.
\end{theorem}

\begin{IEEEproof}
From \eqref{eqn-18}, \eqref{eqn-19} and \eqref{eqn-22}, we have
\begin{align*}
D^{+}\mathcal{W}(t)&\leq\ep[U(\phi)(\mathbf{H}(\phi)\mathbf{J}_{2}(\phi)\mathbf{H}^{\top}(\phi)+\mathbf{L}(\phi)+\mathbf{G}(\phi)u)]  \nonumber  \\
&=-\ep\left[\frac{\mathsf{K}U^{2}(\phi)}{\|U(\phi)\|+\varpi}\right]=:-\mathsf{K}\eta(\phi),
\end{align*}
where $\eta(\phi):=\ep[\frac{U^{2}(\phi)}{\|U(\phi)\|+\varpi}]$. From \cite[Thm. 3.1]{Pepe2006lyapunov}, the functional $\mathcal{W}(t)$ converges to the origin with the increase of time, which implies that the stabilization objective is satisfied. In addition, from \eqref{eqn-21} and the manifold invariant condition, the sliding surface is in the safe set. From \eqref{eqn-20}, we have $\ep[|U(\phi(\theta))|]\geq\ep[|U(\xi(\theta))|]$ for all $\phi\in\partial\mathbb{S}$ and $\theta\in[-\Delta, 0]$. From the convergence of the functional $\mathcal{W}(t)$, the state trajectory starting from the initial condition is convergent along the sliding surface, while avoiding to cross the boundary of the safe set. Hence, the safety objective is guaranteed.
\end{IEEEproof}

In Theorem \ref{thm-3}, the sigmoid function is introduced in \eqref{eqn-22} to guarantee the continuity of the controller \eqref{eqn-19}; see \cite[Sec. 1.2.1]{Shtessel2014sliding} for more discussion. The conditions \eqref{eqn-20}-\eqref{eqn-21} are for the boundary of the safe set, and can be strengthened to the case when the state trajectory approaches to the boundary $\partial\mathbb{S}$. More precisely, let $\mathbb{S}_{1}:=\mathbb{S}-\mathcal{PC}([-\Delta, 0], \mathbf{B}(\varpi))$ with sufficiently small $\varpi>0$, and the conditions \eqref{eqn-20}-\eqref{eqn-21} are replaced to
\begin{align*}
&\ep[U^{2}(\phi)]\geq\ep[U^{2}(\xi)], \quad \forall \phi\in\partial\mathbb{S}_{1}, \\
&\mathbb{A}=\{\phi\in\mathbb{S}: U(\phi)=0\}\subset\mathbb{S}_{1}.
\end{align*}
This strengthened version can be applied to reduce the effects of chattering phenomena on the state trajectory. A specific construction of the functional $U$ is presented below.

\begin{proposition}
\label{prop-1}
Consider the system \eqref{eqn-1} with the safe set $\mathbb{S}$ defined in \eqref{eqn-2}-\eqref{eqn-4} and the initial state $\xi\in\inte(\mathbb{S})$. Let ${V}(\phi)$ be the SCLKF and ${B}(\phi)$ be the SCBKF. If the sliding surface functional is constructed as $U(\phi):=\alpha({V}(\phi))+\beta({B}(\phi))$ with the functions $\alpha, \beta: \mathbb{R}\rightarrow\mathbb{R}^{+}$ satisfying
\begin{equation}
\label{eqn-23}
\ep[U^{2}(\phi)]\geq\ep[U^{2}(\xi)], \quad \forall \phi\in\partial\mathbb{S},
\end{equation}
then the controller \eqref{eqn-19} guarantees simultaneously the stabilization and safety of the system \eqref{eqn-1}.
\end{proposition}

\begin{IEEEproof}
From Theorem \ref{thm-3}, the stabilization objective is guaranteed via the controller \eqref{eqn-19}, and we next show the satisfaction of the safety objective. From the construction of $U(\phi)$, $U(x_{t})\geq0$ for all $t>0$. Since $D^{+}\mathcal{W}(t)<0$, we have $\ep[U^{2}(x_{t})]<\ep[U^{2}(x_{0})]$ for all $t>0$. From \eqref{eqn-23}, $U(\phi)\geq U(\xi)$ for all $\phi\in\partial\mathbb{S}$, which implies that the state trajectory cannot reach the boundary of the safe set. Note that $\xi\in\inte(\mathbb{S})$, and thus the state trajectory will stay in the safe set, which ensures the safety objective.
\end{IEEEproof}

From Theorem \ref{thm-3} and Proposition \ref{prop-1}, we can see that the controller \eqref{eqn-19} with \eqref{eqn-22} ensures the safe stabilization of the system \eqref{eqn-1}, whereas the safety objective depends on the sliding surface functional satisfying \eqref{eqn-20}-\eqref{eqn-21} or \eqref{eqn-23}, which can be treated as the constraints on the construction of the functional $U(\phi)$ and further offers flexibility for the application of the proposed approach to different dynamical systems.

\section{Numerical Example}
\label{sec-example}

\begin{figure}
\centering
\begin{tabular}{cc}
\hspace{-8pt}\includegraphics[width=0.52\columnwidth]{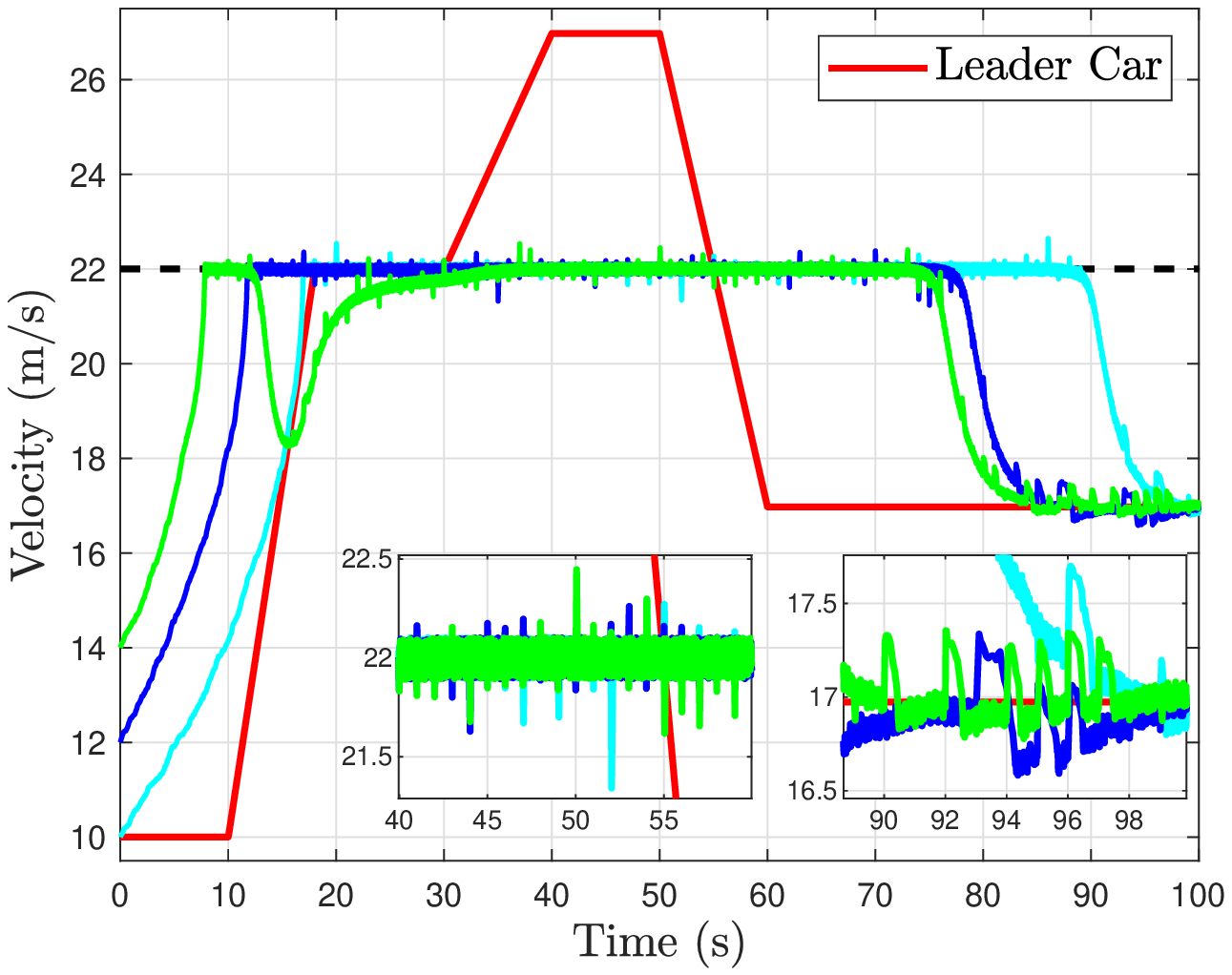} &
\hspace{-15pt}\includegraphics[width=0.52\columnwidth]{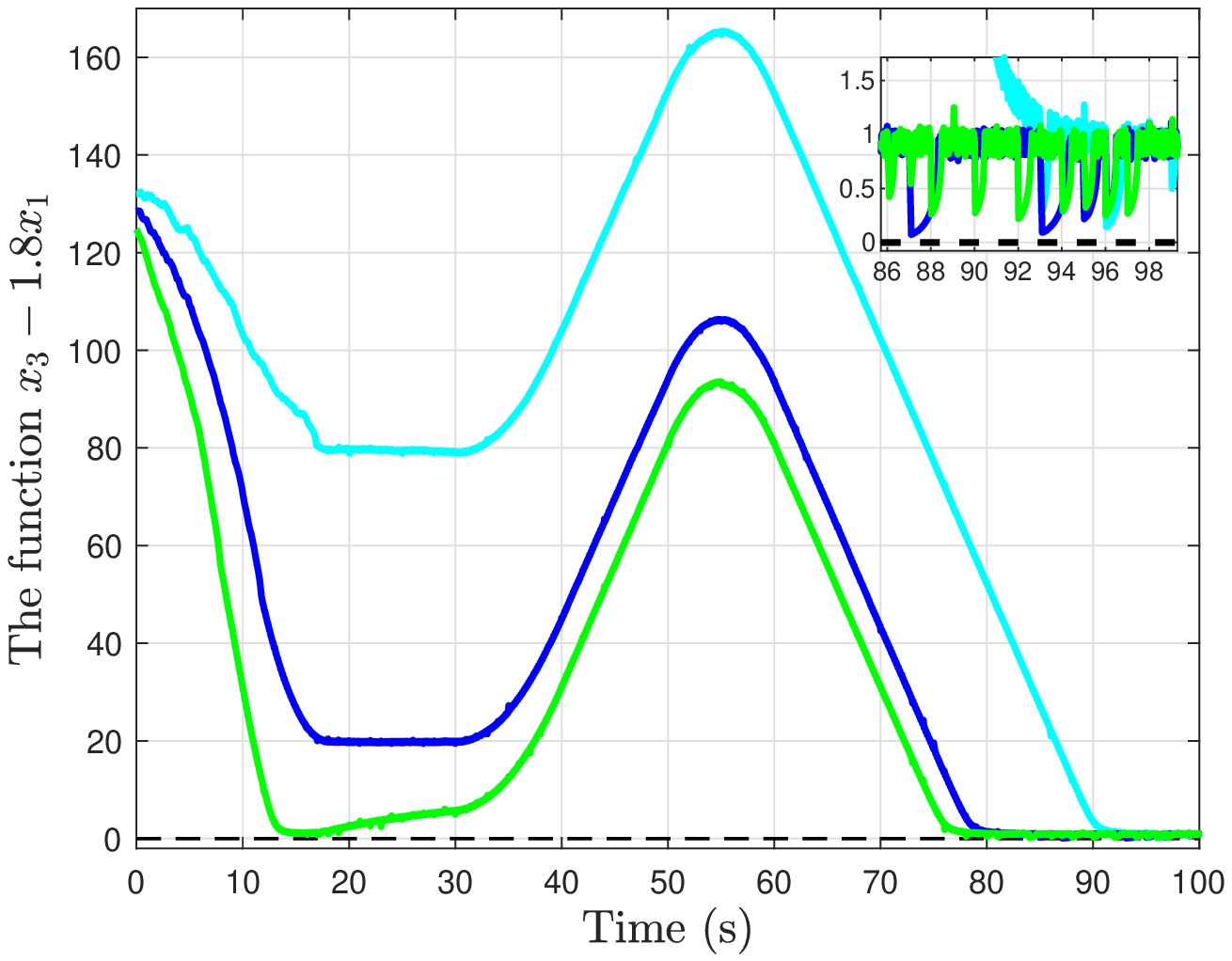} \vspace{-5pt}\\
\hspace{-8pt}\includegraphics[width=0.52\columnwidth]{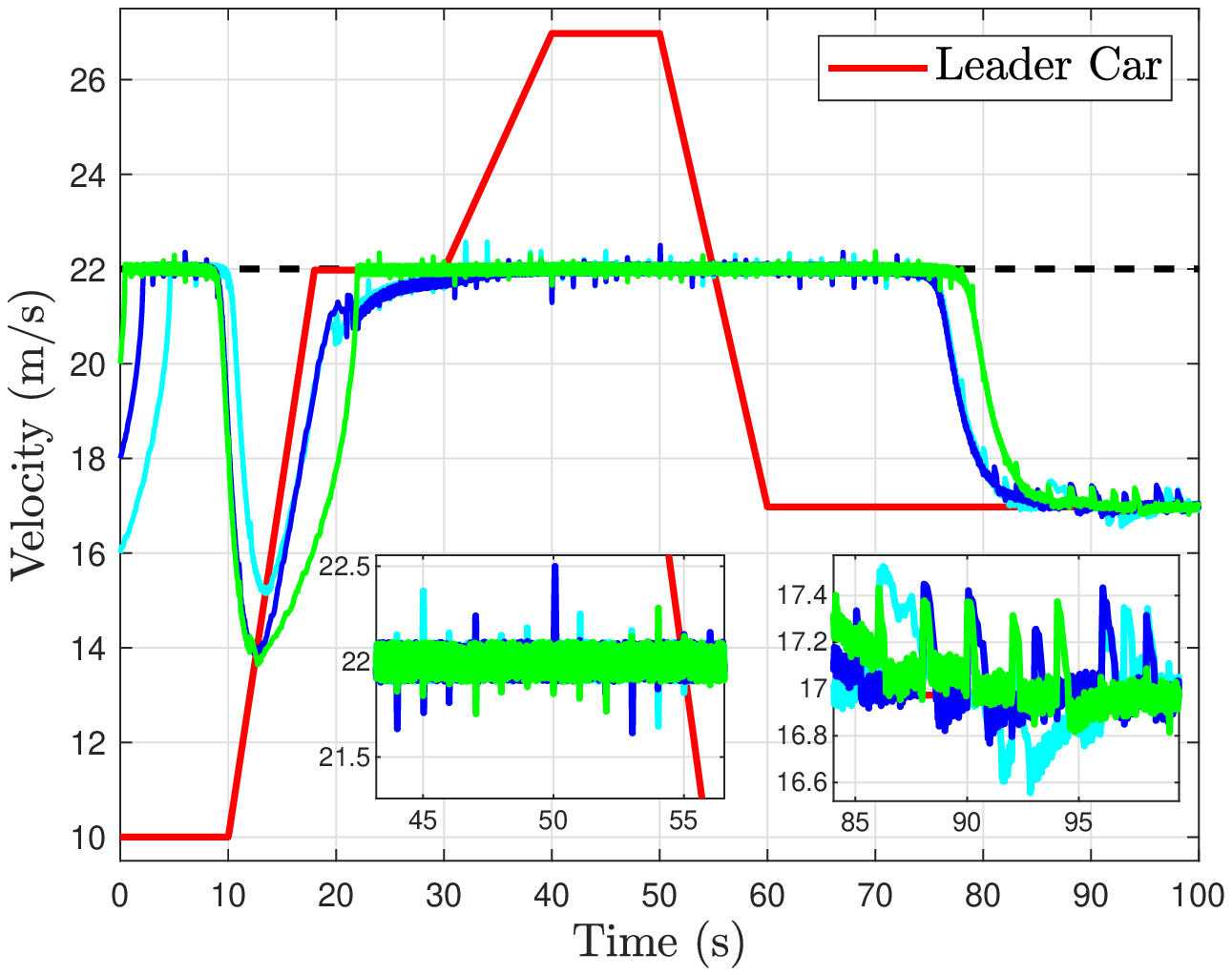} &
\hspace{-15pt}\includegraphics[width=0.52\columnwidth]{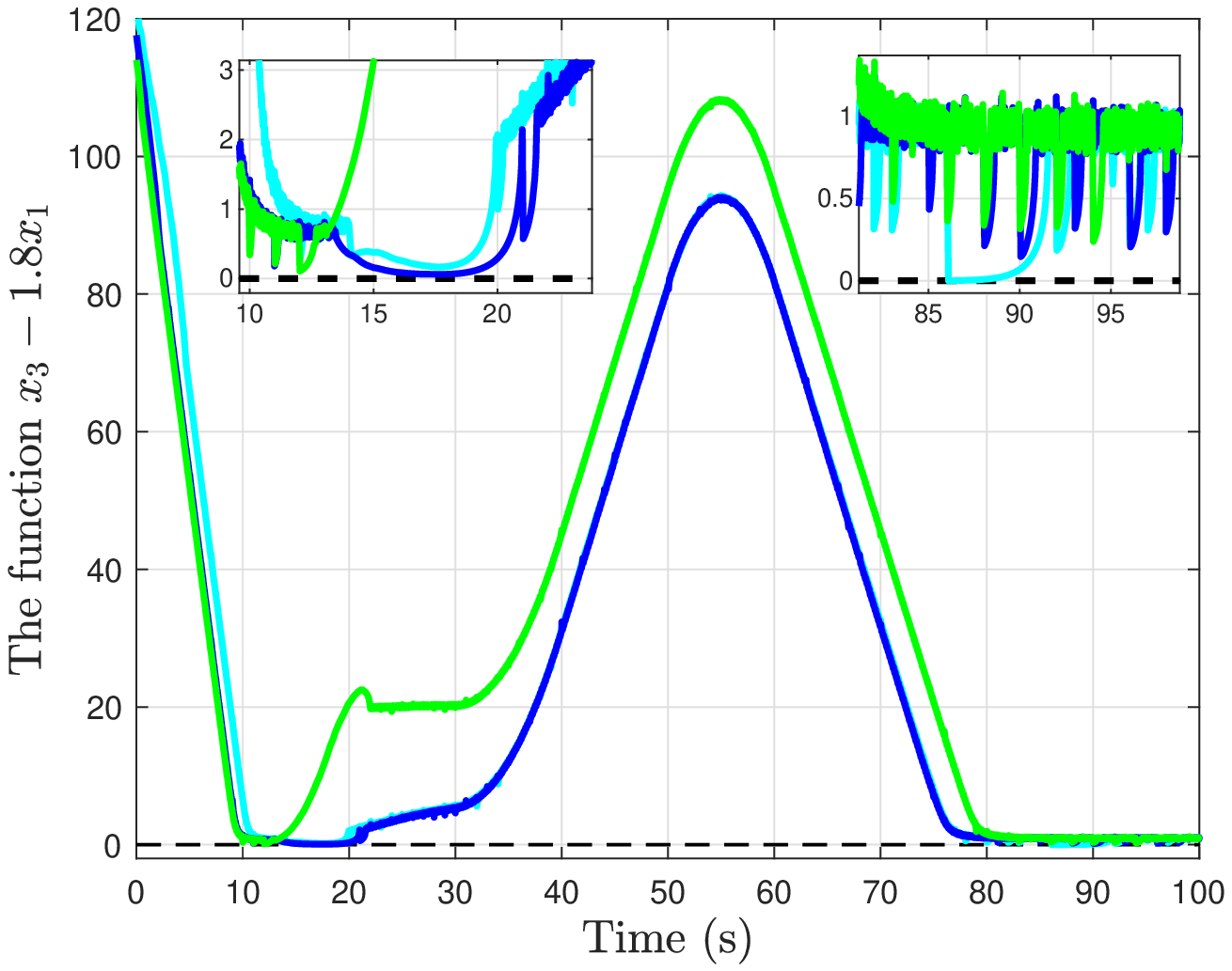} \\
\end{tabular}
\caption{The simulation results under different $\xi=(8+2l, 10, 150)$, $l\in\{1, \ldots, 6\}$. (\textbf{Left}) The velocities of the leader and follower cars. The top sub-figure is for $l\in\{1, 2, 3\}$, and the bottom sub-figure is for $l\in\{4, 5, 6\}$. The red line is the velocity of the leader car, and the black dashed line is the desired velocity $\mathbf{v}_{\mathsf{d}}=22$ of the follower car. (\textbf{Right}) The evolution of the function $x_{3}-1.8x_{1}$ in corresponding cases.}
\label{fig-1}
\end{figure}

Consider the continuous-time car-following model, where both leader and follower cars are modeled as a point-masses and are assumed to move in a straight line \cite{Ames2016control}. Due to the delayed reactions from the driver of the follower car and the external disturbance caused by the road or wind, the car-following model is given as the stochastic time-delay system:
\begin{align}
\label{eqn-24}
\dot{x}(t)=\begin{bmatrix}F(x_{t}(\theta))-F(x(t)) \\ a \\ x_{2}(t)-x_{1}(t)\end{bmatrix}+\begin{bmatrix} u \\ 0 \\ 0\end{bmatrix}+\rho(x_{t})dw(t),
\end{align}
where $x=(x_{1}, x_{2}, x_{3})\in\mathbb{R}^{3}$. $x_{1}\in\mathbb{R}$ and $x_{2}\in\mathbb{R}$ are respectively the velocities of the follower and leader cars (in $\mathrm{m/s}$), $x_{3}\in\mathbb{R}$ is the distance between these two cars (in $\mathrm{m}$). $x_{t}(\theta)=(x_{1t}(\theta), x_{2t}(\theta), x_{3t}(\theta))$, where $\theta\in[-\Delta, 0]$ and $\Delta$ is the upper bound of time delays. $u\in\mathbb{R}$ is the wheel force to be designed as the control input of the follower car. In \eqref{eqn-24}, $F\in\mathbb{R}$ is the total sum of the nonlinear dynamics of car, drag, frictions and disturbances, and $a\in\mathbb{R}$ is the acceleration of the leader car (in $\mathrm{m/s^{2}}$). Following the delay-free case in \cite{Ames2016control}, $F(x)=(a_{0}+a_{1}x_{1}+a_{2}x^{2}_{1})/M$ with the mass $M>0$ of the follower car (in $\mathrm{kg}$) and constants $a_{0}, a_{1}, a_{2}\in\mathbb{R}$ determined empirically. From \cite{Hsia1990robot}, $F(x_{t}(\theta))=(a_{0}+a_{1}x_{1t}(\theta)+a_{2}x^{2}_{1t}(\theta))/M$ is used in \eqref{eqn-24} to estimate the function $F(x(t))$. Finally, $\rho(x_{t})$ is continuous and $w(t)\in\mathbb{R}$ is a one-dimensional Wiener noise.

\begin{figure}
\centering
\begin{tabular}{cc}
\hspace{-8pt}\includegraphics[width=0.52\columnwidth]{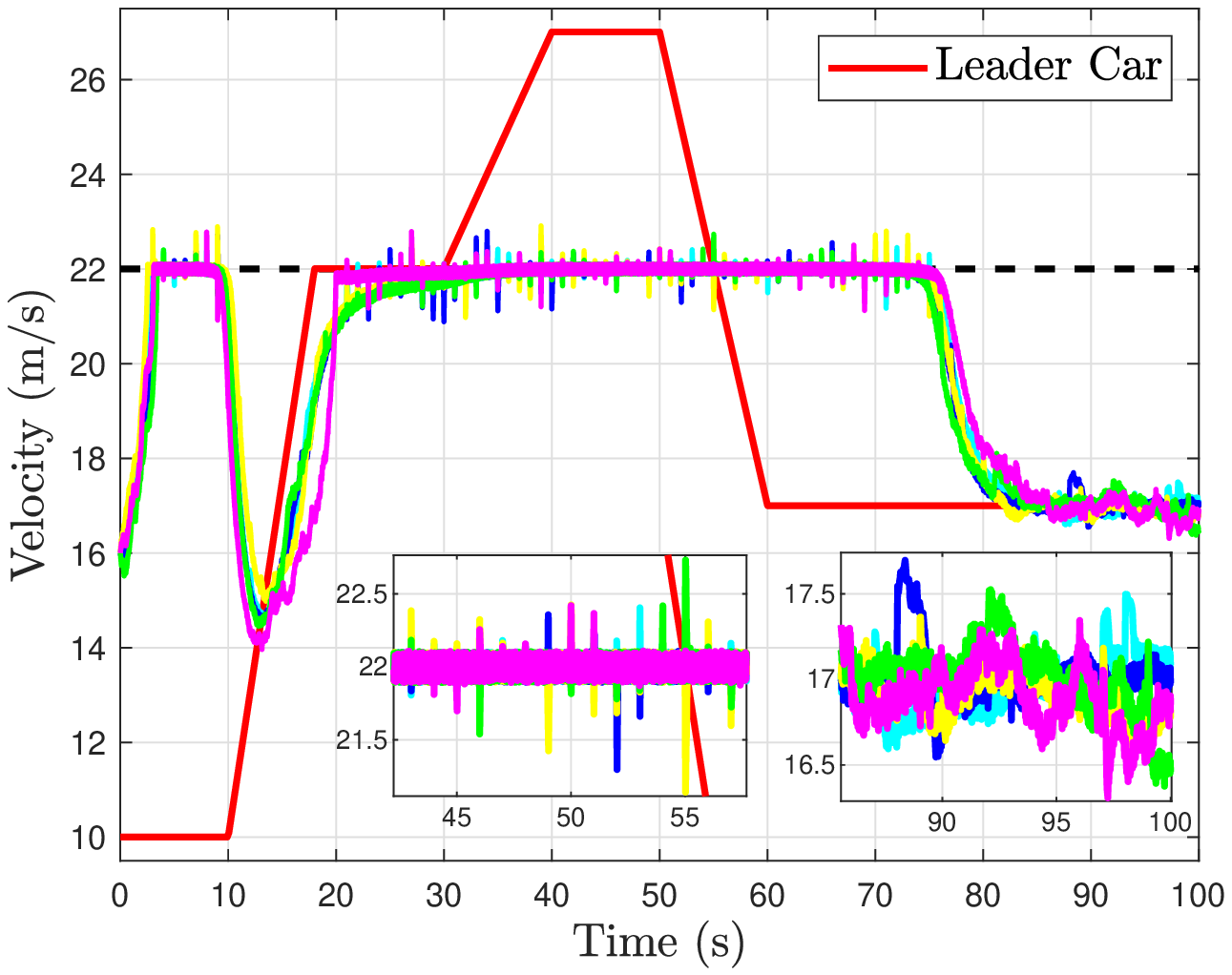} &
\hspace{-15pt}\includegraphics[width=0.52\columnwidth]{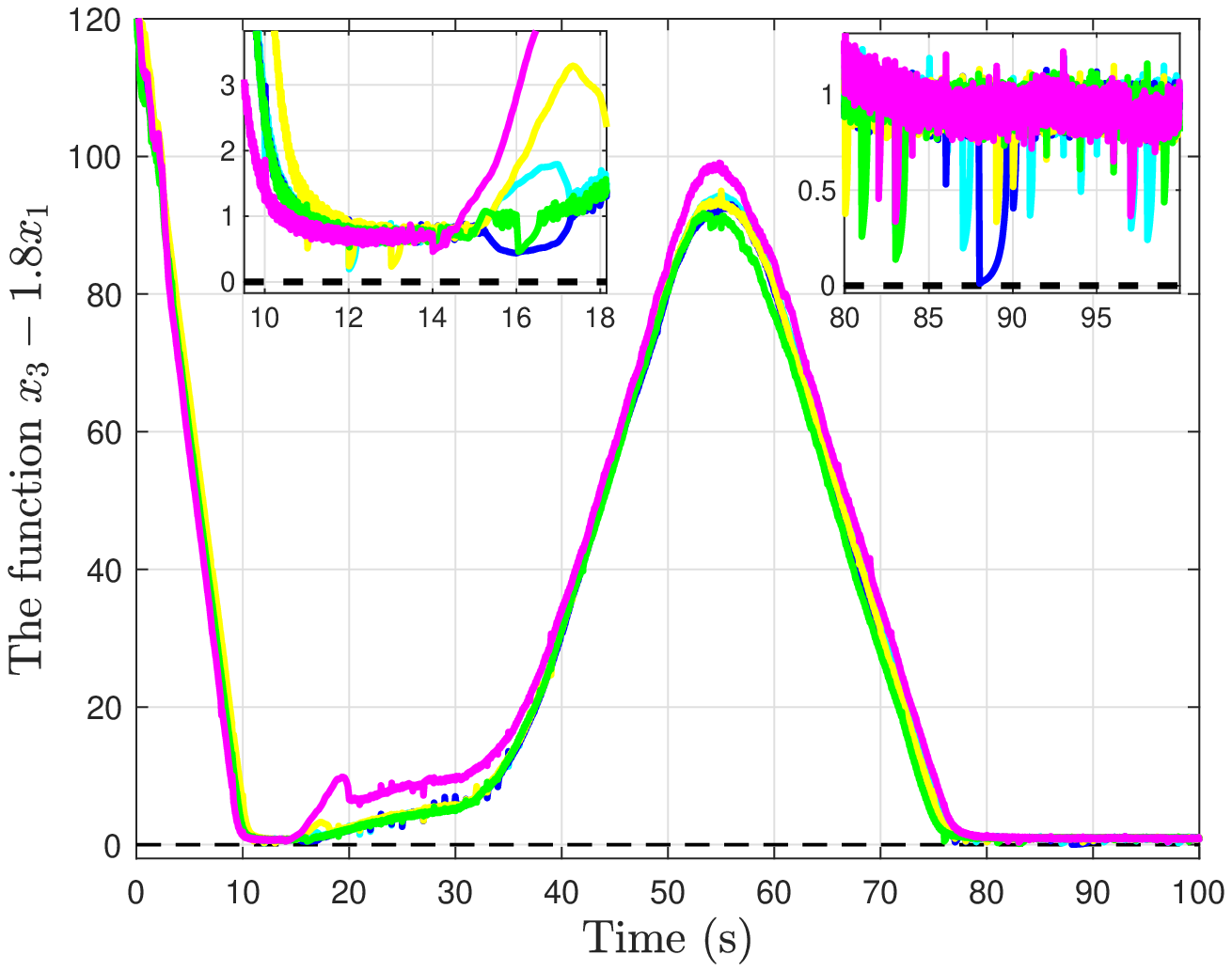} \vspace{-5pt} \\
\hspace{-8pt}\includegraphics[width=0.52\columnwidth]{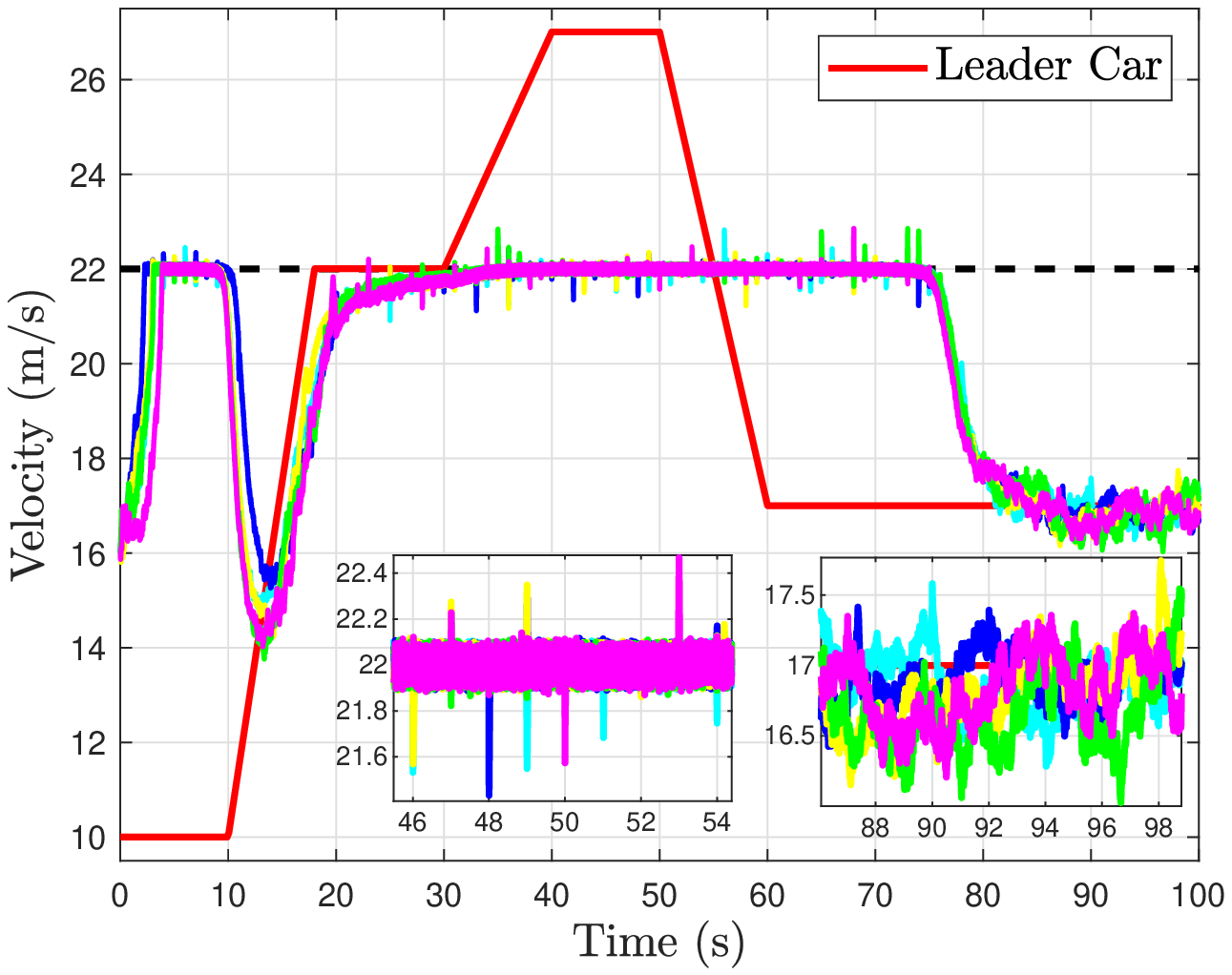} &
\hspace{-15pt}\includegraphics[width=0.52\columnwidth]{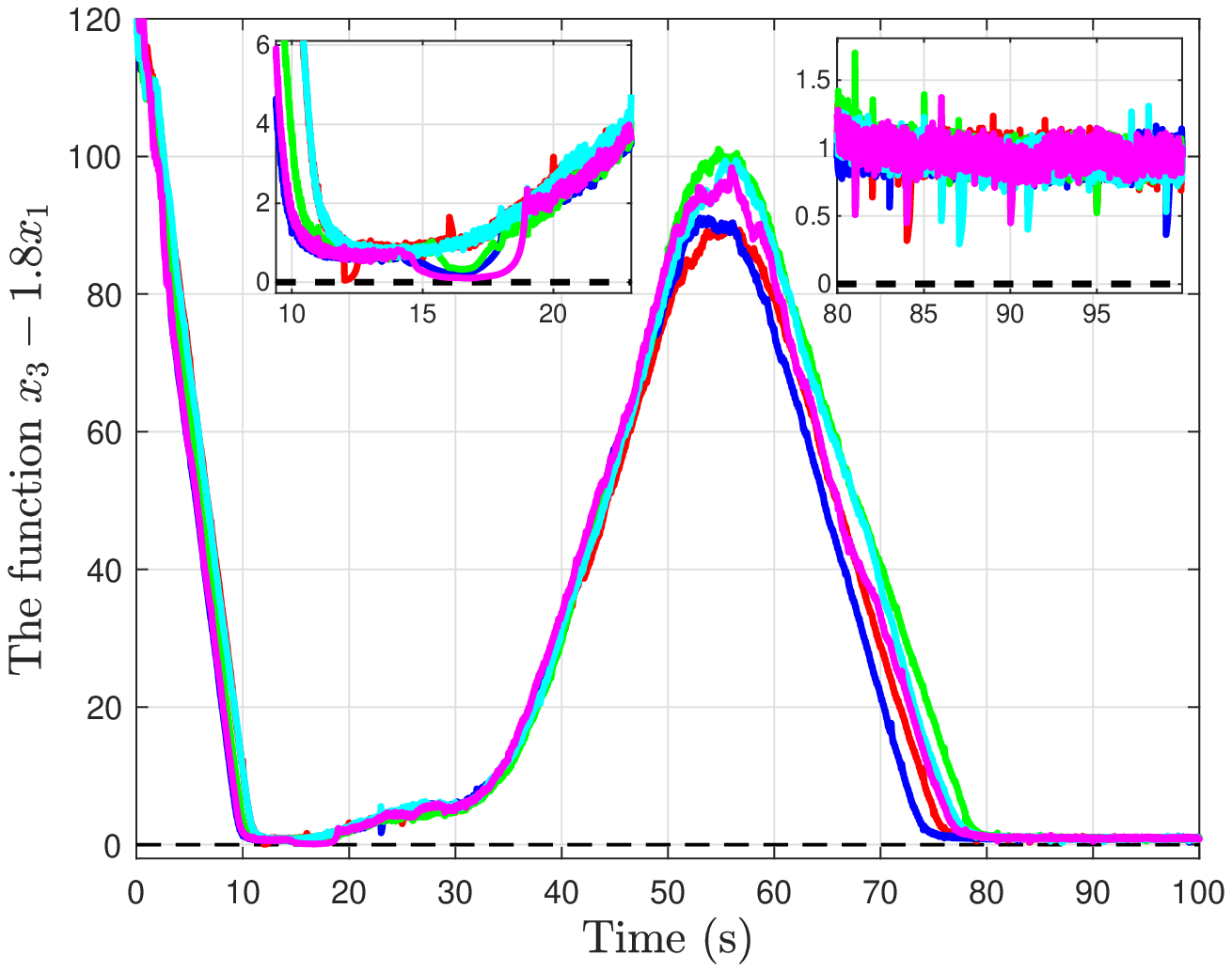}
\end{tabular}
\caption{The simulation results under different $\rho(\phi)=\ell(\phi_{1}, 0, \phi_{3})$, $\ell\in\{1, \ldots, 10\}$. (\textbf{Left}) The velocities of the leader and follower cars. The top sub-figure is for $\ell\in\{1, \ldots, 5\}$, and the bottom sub-figure is for $l\in\{6, \ldots, 10\}$. The red line is the velocity of the leader car, and the black dashed line is the desired velocity $\mathbf{v}_{\mathsf{d}}=22$ of the follower car. (\textbf{Right}) The evolution of the function $x_{3}-1.8x_{1}$ in corresponding cases.}
\label{fig-2}
\end{figure}

In the car-following model, the follower car is expected to follow the leader car in a desired velocity while avoiding to collide with the leader car. To achieve the desired velocity, which is denoted as $\mathbf{v}_{\mathsf{d}}\in\mathbb{R}$, we introduce the following CLKF
\begin{align}
\label{eqn-25}
V(\phi):=(\phi_{1}(0)-\mathbf{v}_{\mathsf{d}})^{2}+\int^{0}_{-\Delta}(\phi_{1}(\tau)-\mathbf{v}_{\mathsf{d}})^{2}d\tau,
\end{align}
where $\phi(\theta)=(\phi_{1}(\theta), \phi_{2}(\theta), \phi_{3}(\theta))=x_{t}(\theta)$. Therefore, to achieve the desired velocity is equivalent to guarantee the convergence of the functional $V(\phi)$, which involves the design of the stabilizing controller. To avoid the collision between the follower and leader cars, their distance needs to be always nonnegative and we introduce the following functional
\begin{equation*}
h(\phi)=\phi_{3}(0)-\mathbf{t}\phi_{1}(0)-0.01\int^{0}_{-\Delta}(\phi_{3}(\tau)-\mathbf{t}\phi_{1}(\tau))^{2}d\tau,
\end{equation*}
where $\mathbf{t}:=1.8$s is the desired time headway (see also \cite{Ames2016control}). Therefore, the collision avoidance is guaranteed when $h(\phi)\geq0$, which further implies $\phi_{3}(0)-1.8\phi_{1}(0)\geq0$. With the functional $h(\phi)$, the CBKF is defined as
\begin{align}
\label{eqn-26}
B(\phi)=\ln(1+1/h(\phi)),
\end{align}
With the functionals \eqref{eqn-25}-\eqref{eqn-26}, we define the sliding surface functional as $U(\phi):=V(\phi)+\varrho B(\phi)$ with $\varrho>0$ to ensure \eqref{eqn-23}. In particular, if $\phi\rightarrow\partial\mathbb{S}$, then $h(\phi)\rightarrow0$ and $B(\phi)\rightarrow+\infty$, which hence shows the existence of $\varrho>0$.

Let $M=1650, a_{0}=0.1, a_{1}=5, a_{2}=0.25, \mathbf{v}_{\mathsf{d}}=22, \varrho=50, \Delta=0.2$ and $a\in[-2.5, 2.5]$. From the controller \eqref{eqn-19}, we choose $\mathbf{K}(x_{t})=\mathsf{K}U(x_{t})/(\|U(x_{t})\|+0.1)$ with $\mathsf{K}>0$. In the following, we consider two cases. The first case is different initial states while the fixed functional $\rho$, that is, $\xi=(8+2l, 10, 150)$, $l\in\{1, \ldots, 6\}$ and $\rho(\phi):=0.05(\phi_{1}, 0, \phi_{3})$. In this case, we choose $\mathsf{K}=10$, and the simulation results are shown in Fig. \ref{fig-1}, which shows the velocity evolution of the follower car and the function $\phi_{3}(0)-1.8\phi_{1}(0)$ under different initial states. From Fig. \ref{fig-1}, $\phi_{3}(0)-1.8\phi_{1}(0)\geq0$, that is, $x_{3}(t)-1.8x_{1}(t)\geq0$ for all $t\geq0$. Hence, the distance between the follower and leader cars are positive, and thus the safety is achieved under different initial states. The second case is the fixed initial state while different values of the functional $\rho$. In this case, let $\xi=(16, 10, 150)$, $\rho(\phi):=\ell(\phi_{1}, 0, \phi_{3})$, $\ell\in\{1, \ldots, 10\}$, and $\mathsf{K}=15$. The simulation results are shown in Fig. \ref{fig-2}, which implies the satisfaction of the safe stabilization. From Figs. \ref{fig-1}-\ref{fig-2}, we can see clearly the chattering phenomena caused by the sliding mode based control design and the effects of the random disturbance on the state trajectory, which results in a potential future topic on how to reduce these effects. However, due to the explicit representation of the controller \eqref{eqn-19}, we can improve the sliding surface functional or adjust the constant $\mathsf{K}$ in \eqref{eqn-19} to guarantee the safe stabilization in different cases.

\section{Conclusion}
\label{sec-conclusion}

This paper provided a framework for the control design of stochastic time-delay systems. Both control Lyapunov-Krasovskii and barrier-Krasovskii functionals were proposed to investigate the stabilization and safety control problems individually. To achieve the safety and stabilization objectives simultaneously, the proposed Krasovskii-type control functionals were combined together such that the stabilizing and safety controllers can be merged. Future work will incorporate input constraints in the proposed approach and extend the proposed approach to the distributed case.


\end{document}